\documentclass[%
 aip,
 amsmath,amssymb,
 reprint,%
]{revtex4-2}

\usepackage{graphicx}
\usepackage{dcolumn}
\usepackage{bm}
\usepackage[utf8]{inputenc}
\usepackage[T1]{fontenc}
\usepackage{mathptmx}
\usepackage{etoolbox}
\usepackage{amsmath}
\usepackage{amssymb}
\usepackage{xcolor}
\usepackage{float}
\usepackage{afterpage}
\usepackage{capt-of}

\newcommand{\be}{\begin{equation}}
\newcommand{\ee}{\end{equation}}
\newcommand{\bea}{\begin{eqnarray}}
\newcommand{\eea}{\end{eqnarray}}

\newcommand{\la}{\langle}
\newcommand{\ra}{\rangle}

\newcommand{\eps}{\varepsilon}

\makeatletter
\def\@email#1#2{%
 \endgroup
 \patchcmd{\titleblock@produce}
  {\frontmatter@RRAPformat}
  {\frontmatter@RRAPformat{\produce@RRAP{*#1\href{mailto:#2}{#2}}}\frontmatter@RRAPformat}
  {}{}
}%
\makeatother
\begin{document}

\preprint{AIP/123-QED}

\title[]{Chaotic Switching In The Minimal Pendula Network}
\author{P. Ebrahimzadeh}
 \affiliation{PGI-14, Forschungszentrum Juelich, Germany}
 \email{p.ebrahimzadeh@fz-juelich.de}
\author{M. Schiek}%
 \affiliation{PGI-14, Forschungszentrum Juelich, Germany}
\author{Y. Maistrenko}
 \affiliation{Institute of Mathematics and Technical Centre, NAS of Ukraine, Tereschenkivska Str. 3, 01030 Kyiv, Ukraine}

\date{\today}

\begin{abstract}
We report the chaotic switching phenomenon in the minimal $N = 3$ pendula network with global coupling. Analyzing the stability conditions of the chimera states and their dependence on the parameters, three scenarios of chaotic switchings are identified: 1) a riddling bifurcation scenario, where an unstable periodic orbit inside the chimera manifold becomes transversally unstable, 2) a blowout bifurcation scenario, where the switching is caused by the transverse destabilization of the chaotic chimera with respect to its manifold, and 3) switchings between "laminar" saddle chimeras within a global "turbulent" attractor. The results are obtained based on the detailed examination of the existing regimes including chimera states, limit cycles and fixed points, their multistability and switching regime.
In the parameter regions where the chaotic chimeras coexist with stable non-chaotic solutions, the switching trajectory can eventually escape to a stable solution, causing an additional unpredictability in the system behavior, as it is difficult to predict the escaping moment.
\end{abstract}

\maketitle

\begin{quotation}
After their discovery two decades ago, chimera states were generally referred to spatio-temporal patterns consisting of coherent and incoherent groups of oscillators emerging in networks of identical oscillators with non-local coupling. Further studies identified a variety of chimeras in networks with different types of coupling. One of them is the so-called mixed-mode chimera state which is comprised of one group of oscillators in quiescence and the other rotating with a non-zero average frequency. These mixed-mode chimeras were originally observed in pendula networks and are found in adaptive networks as well as in neuronal models. In neuroscience, the mixed-mode chimeras may be considered as images of bump states, where the coexistence of the firing neurons with the quiescent ones are a prospect for modeling working memory in the brain.
The switching dynamics between the mixed-mode chimeras arise as progressive events where the system trajectory alternates between different chimera replicas. In pendula networks, we show that spontaneous switching between the chaotic chimera states arises due to loss of transversal stability, transforming them into chaotic saddles. 
\end{quotation}

\section{\label{intro}Introduction}

The discovery of chimera states at the edge of millennium in 2002 \cite{kuramoto2002coexistence, PhysRevLett.93.174102} has opened a new line of research in network dynamics theory and application \cite{davidsen2024introduction}. It continues to deliver novel fascinating phenomena both in theory and practice (see the review papers \cite{Panaggio_2015, scholl2016synchronization, omel2019chimerapedia, zakharova2020chimera, mishra2023chimeras} and references therein). Originally, chimeras were reported as spatio-temporal patterns in large networks of identical non-locally (and non-globally) coupled phase oscillators, in which one group of oscillators is synchronized and the other group behaves asynchronously.

The notion of the weak chimera state \cite{ashwin2015weak} marked an important milestone in the study of chimera states, particularly in small networks. To this end, aside from the main frequency-synchronized group, there exists at least one oscillator with a different average frequency, also known as {\it solitary state} \cite{jaros2018solitary}. The simplest non-trivial situation of this kind arises for $N=3$ coupled identical oscillators, when two oscillators are frequency synchronized but the third one is rotating with a different average frequency \cite{maistrenko2017smallest}.
It was found that, in the pendula networks different chimera types can arise due to the pendulum bistability above the Tricomi bifurcation curve \cite{tricomi1933integrazione}. A major contribution to the system dynamics is given by the so-called {\it mixed-mode chimeras} states \cite{ebrahimzadeh2022mixed}, each of which contains both rotating and non-rotating (slighly oscillating) modes. This kind of remarkable behavior was also reported recently for a network of coupled Hodgkin–Huxley neurons \cite{rossi2025transients}.

A significant phenomenon in complex dynamical systems is the coexistence of attractors. In such multistable systems, switching dynamics \cite{rabinovich2001dynamical,ashwin2005instability} is possible such that the system trajectory alternates between different attractors. The mechanisms of switching dynamics include noise-induced attractor hopping or adaptation to external input \cite{kraut2002multistability}. In contrast, switching dynamics can occur spontaneously when the attractors change their stability transforming into saddle states. We refer to Ref. {\cite{ansmann2016self} and references therein for comprehensive description and examples of switching phenomena. For systems inheriting chimera states, the switching dynamics have been reported as {\it chimera switching} \cite{goldschmidt2019blinking, zhang2020critical, brezetsky2021chimera}.

The paper is organized as follows: we introduce the model as a network of $N=3$ coupled pendula with repulsive coupling in Sec.~\ref{md} and present all of its dynamical states, including different types of chimera states. The stability regions and in-manifold dynamics of these states are discussed. In Secs.~\ref{rb}--\ref{lt} three scenarios of the chaotic switching states and bifurcations leading to the transversal destabilization of the chimeras are presented. We conclude our results in Sec.~\ref{csd} along with a discussion of prospective applications.

\section{\label{md} The Model and Its Dynamics}
We consider a network of globally coupled pendula given by the equations
\be
\label{PN}
m\ddot\phi_i + \eps \dot \phi_i + r \sin{\phi_i}= w + \frac{\mu}{N} \sum_{j=1}^{N} \sin[\phi_j - \phi_i - \alpha],
\ee
where $i=1,...,N$, $\alpha$ is the phase-shift, $\mu$ the coupling strength between the pendula, $m$, $\eps$ and $w$ are inertia, damping and eigenfrequency of each pendulum, respectively. 
In the absence of the coupling, the dynamics of a free pendulum consists of a stable fixed point $\phi^O = \sin^{-1} ({w}/{r})$, its corresponding saddle $\phi^{S} = \pi - \sin^{-1} ({w}/{r})$, and a limit cycle with an average rotation velocity $\bar\omega \equiv \la \dot\phi \ra = w/\eps$. 
The limit cycle is born at Tricomi bifurcation curve\cite{tricomi1933integrazione} $T(\eps,w)<1$, and the fixed points vanish in an inverse saddle-node bifurcation at $w/r = 1$. A notable difference between system~(\ref{PN}) and its corresponding phase oscillator model (Kuramoto model without inertia when $m=r=0$) is the existence of the region of bistability bounded by the Tricomi and the saddle-node curves, where both the  stable fixed point and the limit cycle coexist. Without loss of generality, the system parameters are set to $w=0$, $r=1$, and $\eps=0.1$.

In the attractive coupling region $\alpha < \pi/2$, the dynamics of system~(\ref{PN}) include: fixed point and limit cycle on a synchronous manifold, chimera states with all the pendula in rotating mode (named below as "standard" chimera) and mixed-mode chimera states in which a group of pendula is quiescent and other group is rotating \cite{ebrahimzadeh2022mixed}. Both types of standard and mixed-mode chimera states form frequency clusters with dynamics of each cluster on a synchronous sub-manifold. For the minimal network of $N=3$ pendula, three chimeric attractors were identified: standard chimera state with the frequencies symbolically denoted as $(1:2:2)$ such that the fast rotating group includes two pendula, mixed-mode chimera no.1 $(0:1:1)$ where the quiescent group with zero rotation frequency includes one pendulum, and mixed-mode chimera no.2 $(0:0:1)$ with two pendula in the quiescent group. The symbolic notation $(\bar\omega_1: \bar\omega_2: \bar\omega_3)$ denotes the average relative rotation frequency of the pendula. 

In this paper, we study the dynamics of system~(\ref{PN}) in the repulsive coupling region $\alpha>\pi/2$ for the minimal network $N=3$. The dynamical behaviors in this region includes: a standard chimera $(1:1:a), a>1$ where the synchronized pendula are slower than the desynchronized one, the out-of-phase mixed-mode chimera no.2 $(0:0:1)$ where the phase-velocity of the quiescent pendula are out-of-phase, the oscillatory state $(0:0:0)$ in which all the pendula are phase and frequency locked with their dynamics manifested as oscillations around the origin, and a $1D$ ring manifold of asynchronous fixed points. For clarity, we will call $(1:1:a)$ the inverted chimera. The main peculiarity of the dynamics in the repulsive coupling region $\alpha>\pi/2$ is the appearance of the chaotic switching behavior found in large regions of the parameter plane. 

\begin{figure*}
\includegraphics[width=\textwidth]{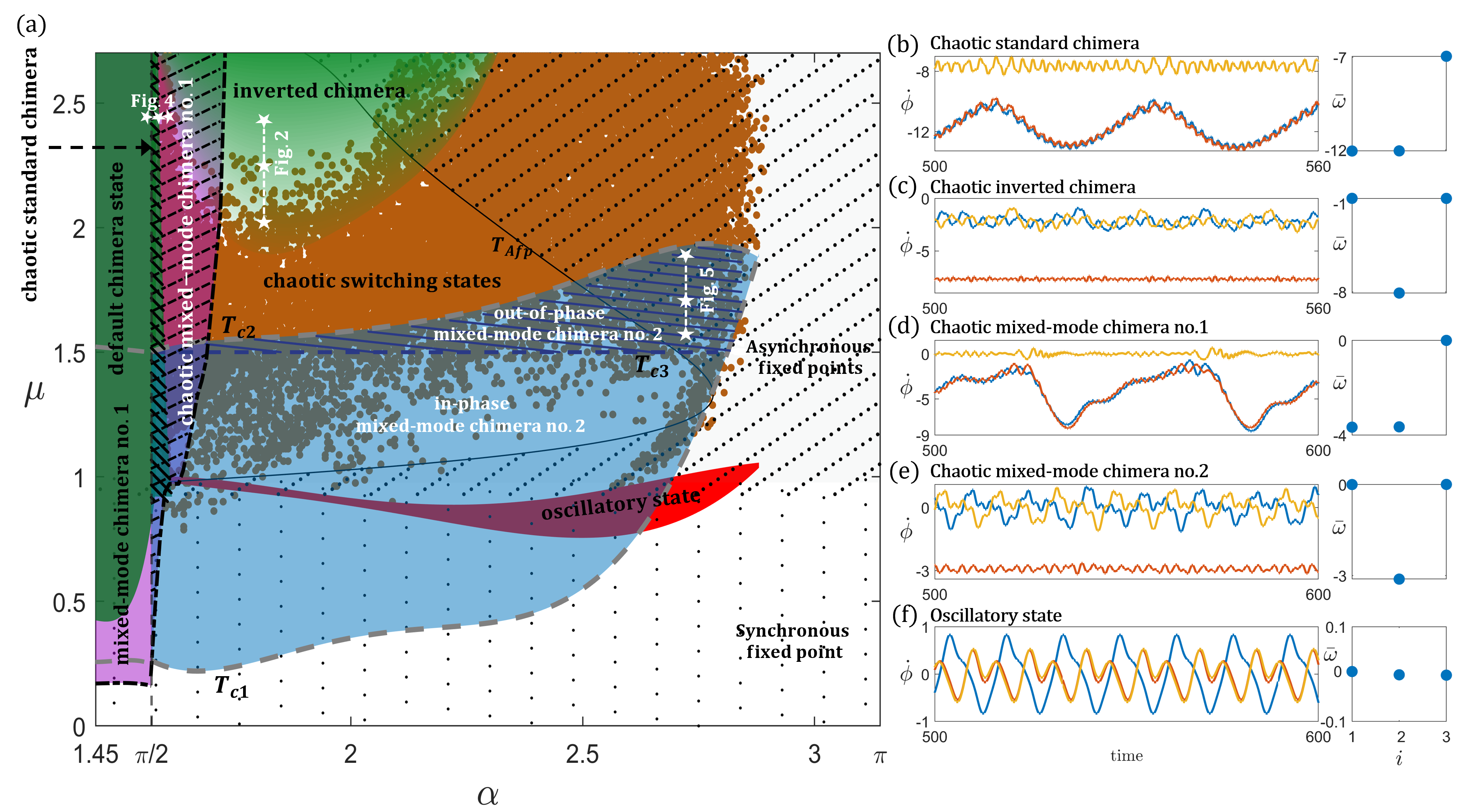}
\caption{\label{fig1}Parameter regions and sample solutions of the system~(\ref{PN}). (a) Stability regions in the $(\alpha, \mu)$-parameter plane including the standard chimera state $(1:2:2)$ in the green region, the mixed-mode chimera no.1 $(0:1:1)$ in purple region, the in-phase mixed-mode chimera no.2 (0:0:1) in the blue region bounded by the bifurcation curves $T_{c2}$ and $T_{c1}$, the out-of-phase mixed-mode chimera no.2 (0:0:1) in the hatched blue region, the chaotic inverted chimera $(1:1:a)$ in the light green region, the oscillatory state where all the pendula have zero mean frequency in the red area, the asynchronous fixed point in the dashed-dotted region beyond curve $T_{Afp}$ and $\mu>1$, the synchronous fixed point in the dotted region for $\mu<1$ and any $\alpha$, and the chaotic switching states in the orange dotted region. The sample solutions include (b) chaotic standard chimera at $(\alpha, \mu) = (1.58, 2.5)$, (c) chaotic inverted chimera at $(\alpha, \mu) = (1.8, 2.5)$, (d) chaotic mixed-mode chimera no.1 at $(\alpha, \mu) = (1.6, 2.0)$, (e) chaotic out-of-phase mixed-mode chimera no.2 at $(\alpha, \mu) = (2.6, 1.55)$ and (f) frequency-locked oscillatory state at $(\alpha, \mu) = (2.6, 1.2)$.}
\end{figure*}
Results of direct numerical simulation of system~(\ref{PN}) for $N=3$ coupled pendula in the two-parameter plane of the phase-shift $\alpha$ and coupling strength $\mu$ are presented in Fig.~\ref{fig1}(left) along with the velocity-time plots of sample solutions and their respective average frequency profiles, Fig.~\ref{fig1}(right).  This figure contains dynamical states of the system (\ref{PN}) including different chimera states and chaotic switchings. See appendix \ref{A1} for simulation details.
In the attractive region $\alpha < \pi/2$, three chimera states are identified \cite{ebrahimzadeh2022mixed}: standard chimera $(1:2:2)$ in dark green, mixed-mode chimera no.1 $(0:1:1)$ in purple (above the dotted-dashed line coexisting with standard chimera $(1:2:2)$) and mixed-mode chimera no.2 $(0:0:1)$ in the blue region confined between the $T_{c1}$ and $T_{c2}$ bifurcation curves.
Increasing the phase-shift beyond $\alpha > \pi/2$, the standard chimera $(1:2:2)$ and mixed-mode no.1 $(0:1:1)$ become chaotic, see appendix \ref{A2}. The existence region of these chaotic attractors are marked with the narrow downwards dashed area for the standard chimera and upwards dashed areas for the mixed-mode chimera no.1. 
Mixed-mode chimera no.2, however, remains non-chaotic for all values of phase-shift $\alpha$ and coupling strength $\mu$ with dynamics of the quiescent pendula remaining on a synchronous sub-manifold. Above $T_{c3}$, notably, a chimera state with the same frequency profile as the mixed-mode no.2 exists, but the dynamics of the quiescent pendula are out-of-phase. These two mixed-mode chimeras coexist in the dashed blue region for $\alpha > \pi / 2$. Moreover, the out-of-phase mixed-mode chimera no.2 becomes chaotic for $\alpha \gtrsim 1.86$.
In the light green region for large values of $\mu$, there exists the inverted chimera in which the fast rotating pendula group reverses to a slower frequency. Beside the chimeric solutions, there are the oscillatory state where all pendula are oscillating around the origin (the red region), the asynchronous ring manifold of fixed points that exist above $\mu=1$ and stabilize for values of $(\alpha, \mu)$ beyond the curve $T_{Afp}$ (dashed-dotted region), and the synchronous fixed points that are stable for all values of $\alpha$ and $\mu<1$ (dotted region). 

Stability of the synchronous fixed point can be obtained analytically. The characteristic equation for the Lyapunov exponents (LEs) of system~(\ref{PN}) with arbitrary size $N$ reads \cite{ebrahimzadeh2022mixed}
\be
\label{CEL}
\Big( z + \sqrt{1 - \mu^2 \sin^2{\alpha}} \Big) \Big( z + \sqrt{1 - \mu^2 \sin^2 {\alpha}} + \mu \cos{\alpha} \Big)^{N-1} = 0,
\ee
where $z =  \lambda^2 + \eps \lambda$. The first term in Eq.~(\ref{CEL}) controls the in-manifold stability and the second term provides two $(N-1)$-multiple LEs for transversal stability. The real part of in-manifold LEs are negative $\Re (\lambda_{\pm}) < 0$ for any $\alpha$ . In the region $\alpha > \pi/2$, however, the transversal LEs acquire positive real value $\Re(\lambda_{+})>0$ for $\mu>1$ and the synchronous fixed point loses its transverse stability.

Our simulations confirm that the major domain of the parameter region in the repulsive regime $\alpha > \pi/2$ for intermediate values of $\mu$ consists of the chaotic switching states, shown by brown dots. The switching behavior is characterized as a progressive sequence of events in which the system trajectory changes from one characteristic state to another or returns back to the original state. For the two-cluster chimera states of system~(\ref{PN}), there exist $C(N, k) = N! / (N-k)!\, k!$ replicas for each chimera imposed by the symmetry of the identical pendula, such that $N$ is the total number of pendula and $k$ is the number of pendula in the cluster. For the minimal pendula network $N=3$, there are three replicas for each chimera type. Each chimera replica reside in its invariant subspace, and the chaotic switching appears after the chimera states become transversally unstable and the system trajectory switches between the chimera replicas. Moreover, the chaotic switching is not restricted to switches between the chimera replicas of the same type, but the trajectory may leave the subspace and fall into the other chimera types as well. In the following we describe three scenarios for the appearance of the chaotic switchings in system~(\ref{PN}).

\section{\label{rb} Riddling Bifurcation Scenario}

The first scenario of chaotic switching involves the riddling bifurcation for the chaotic inverted chimera. The riddling bifurcation describes the mechanism in a chaotic system in which a periodic orbit embedded in the chaotic attractor becomes unstable in direction transversal to its invariant manifold while the whole chaotic state remains stable in average. As a result, tongue structures appear at the transverse direction of the unstable orbit, where the trajectory escapes the attractor \cite{lai1996riddling}.
\begin{figure}
\includegraphics[width=\columnwidth]{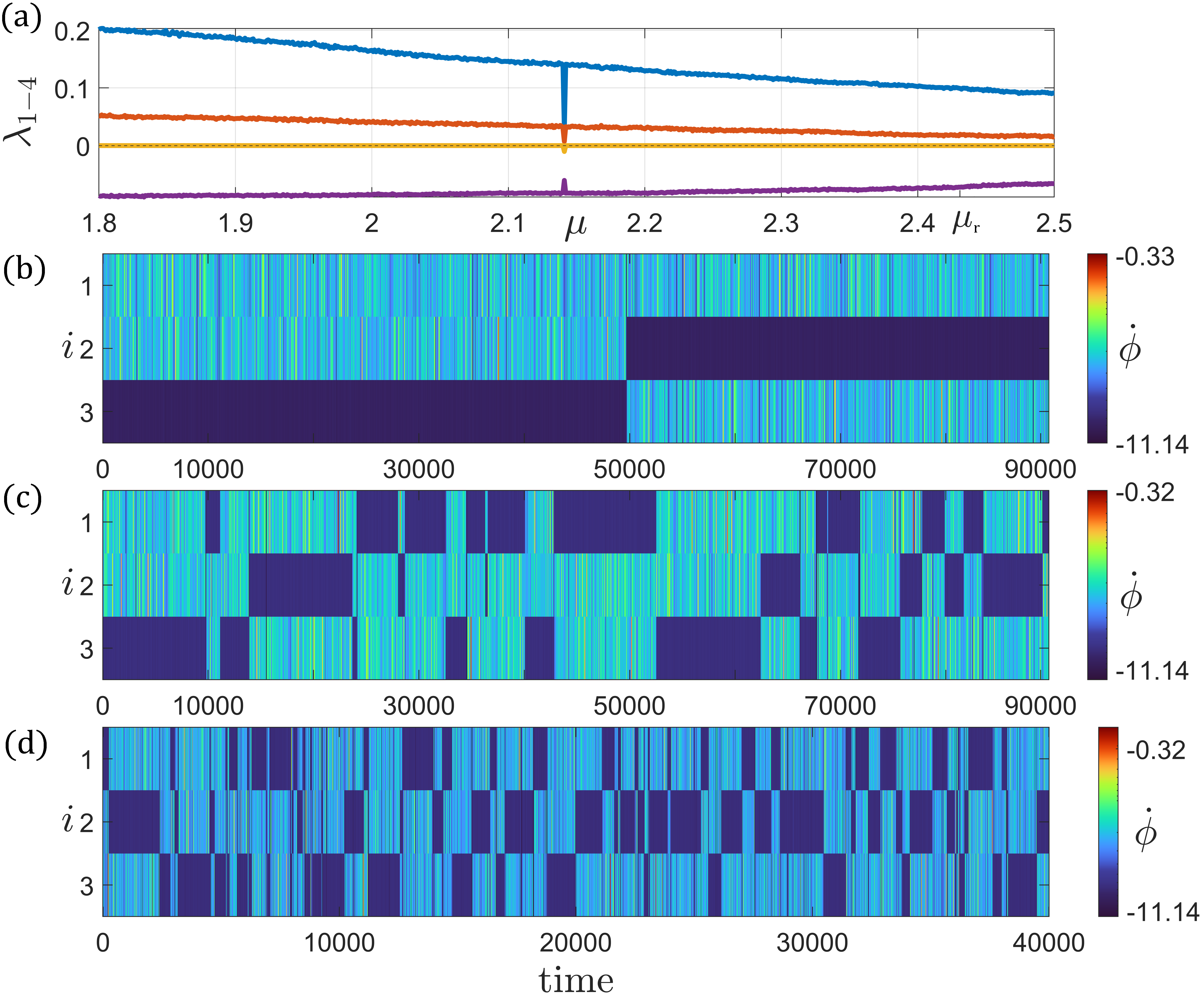}
\includegraphics[width=\columnwidth]{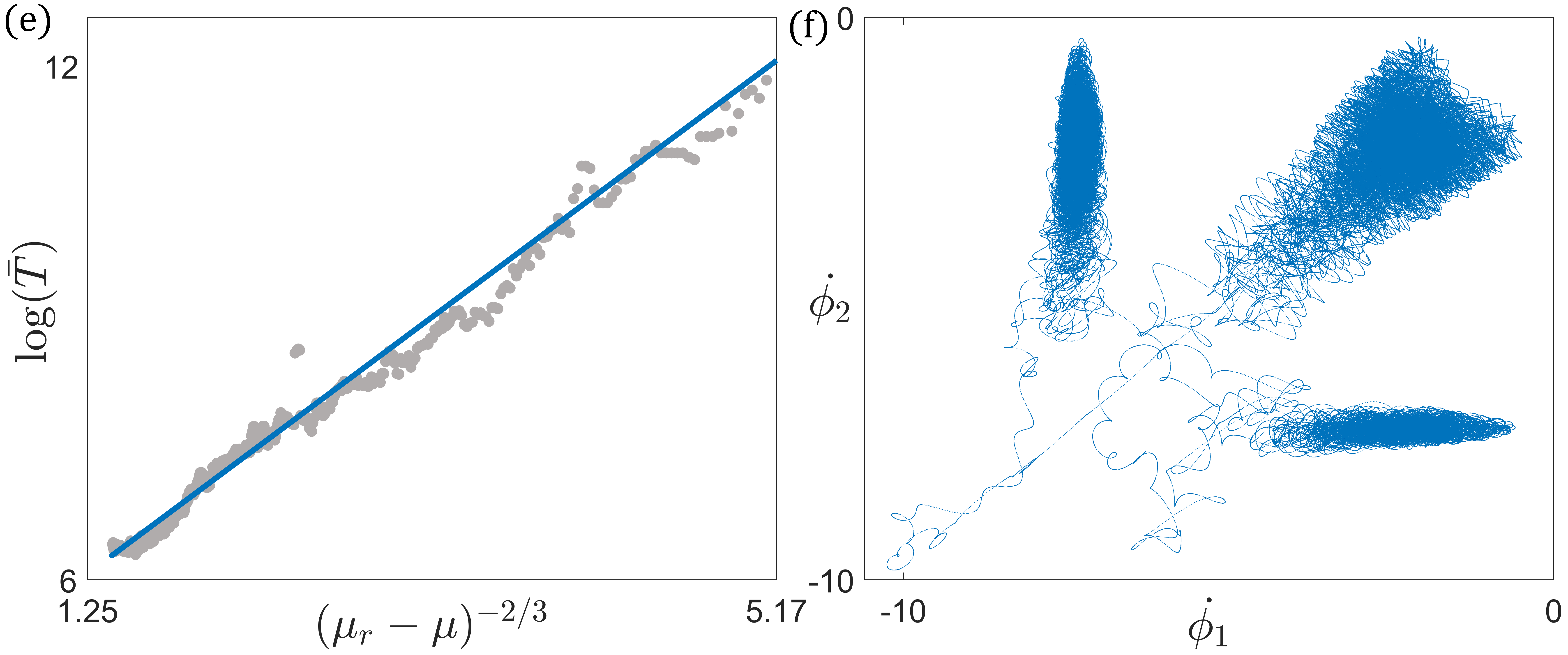}
\caption{ \label{fig2} Riddling bifurcation scenario of the chaotic switching. (a) Lyapunov exponents of the inverted chimera state are calculated for fixed phase-shift $\alpha = 1.8$ and varying coupling strength: the inverted chimera is hyperchaotic with two positive Lyapunov exponents. After the riddling bifurcation at $\mu_r \approx 2.41$, decreasing the coupling strength $\mu$ the switching periods become shorter (b) $\mu_1 = 2.32$, (c) $\mu_2 = 2.25$, and (d) $\mu_3 = 1.9$. (e) the average switching period of  the inverted chimeras plotted against the distance from the riddling bifurcation point. (f) phase-velocity portrait of the three inverted chimeras.}
\end{figure}
\begin{figure*}
\includegraphics[width=\textwidth]{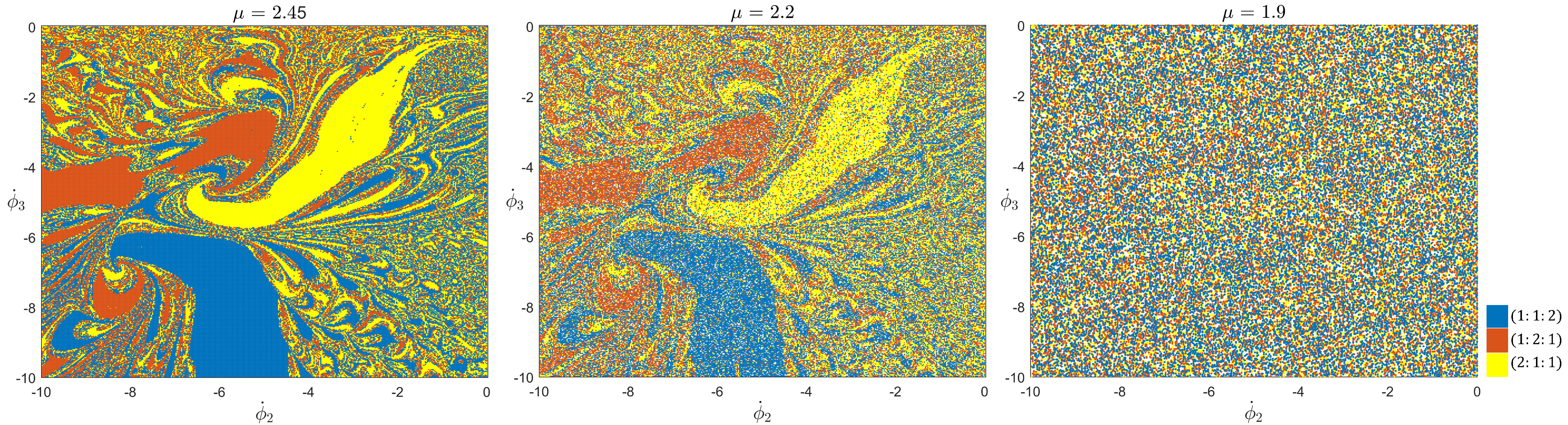}
\caption{\label{fig3} Basins of attraction of the inverted chimera at fixed phase-shift $\alpha = 1.8$. Each color denotes one of the replicas, see the legend. Before the riddling bifurcation at $\mu_r \approx 2.41$, each chimera replica has large basins of attraction with fractal basin boundary $\mu = 2.45$. After the riddling bifurcation for the coupling strength $\mu = 2.2$, the basin of each chimera replica is riddled with the points belonging to the other replicas. Further decrease of coupling strength $\mu = 1.9$, the basins of attraction are completely riddled.}
\end{figure*}
A typical scenario of the riddling-provoked switching in the system~(\ref{PN}) with $N=3$ is illustrated in Fig.~\ref{fig2}, where the phase-shift is fixed at $\alpha=1.8$ and the coupling strength is varied from $\mu = 2.5$ decreasing to $\mu = 1.9$. Four of the Lyapunov exponents of the chaotic inverted chimera are shown in Fig.~\ref{fig2}(a), but it does not demonstrate the characteristic differences of LEs along the parameter line. However, in the three consecutive examples in Fig.~\ref{fig2}(b,c,d) the chaotic switching is observed, first very rare in Fig.~\ref{fig2}(b) and then more often in Fig.~\ref{fig2}(c,d). It is due to the fact that the LEs measure the stability of the chaotic state on average and cannot detect the narrow escaping tongues.
As shown in Fig.~\ref{fig2}(a), the inverted chimera is hyperchaotic with two positive Lyapunov exponents and remains transversally stable in average for large values of $\mu>\mu_r \approx 2.41$ and no switching is observed. The inverted chimera then loses its transverse stability in a riddling bifurcation at $\mu=\mu_r$ and the system trajectory can escape the chimera manifold.
An example of the switching is shown in Fig.~\ref{fig2}(b) for $\mu_1 = 2.32$ where the only switching is observed at time $T \simeq 50,000$ between the inverted chimera replicas $(1:1:a)$ and $(1:a:1)$. With further decrease of the coupling strength, the period of the switching between the chimera replicas reduces, resulting in more switching events, see Fig.~\ref{fig2}(c) for $\mu=2.25$ and (d) for $\mu = 1.9$. 
The characteristic average period of the switching in the riddling bifurcation scales superexponentially with respect to the distance from the bifurcation point \cite{lai1996riddling}. Near the bifurcation, the tongue structure is extremely narrow and only a very small fraction of the points at the tongue can escape the attractor. The probability of the trajectory falling into the tongues increases as the parameter moves away from the bifurcation point. For the inverted chimeras, the average switching period between the chimera replicas coincides with the transient time of the riddling bifurcation since, after escaping one of the replicas, the trajectory is immediately attracted to another replica. Then, the average switching period reads
\be
\label{rT}
T \sim \exp[ K d^{\gamma} ],
\ee
where $d = (\mu_r - \mu)$ is the distance from the riddling bifurcation point, $K$ is a constant relating to the maximum positive Lyapunov exponent of the chaotic attractor and $\gamma = -2/3$. Despite the fact that we were not able to detect the unstable periodic orbit inside the chaotic chimera which becomes transversally unstable in the moment of bifurcation $\mu = \mu_r$, our simulations confirm the superexponential scaling law of the switching periods, see Fig.~\ref{fig2}(e). The riddling bifurcation point $\mu_r$ is extrapolated from the data using an error minimization method with the \texttt{fminsreach} function in MATLAB, and is predicted to be $\mu_r \approx 2.412$. Fig.~\ref{fig2}(d) illustrates the phase-velocity profile of the three inverted chimera replicas at parameters $\alpha = 1.8$ and $\mu = 2.2$.

At the riddling bifurcation, the basin of attraction inherits an open and dense set of "holes". That is, for every open set in the basin of the chaotic attractor, there exist points belonging to the other attractor's basin \cite{alexander1992riddled}. The basins of attraction of the three inverted chimera replicas are shown in Fig.~\ref{fig3} where the colors red, blue, and yellow represent each replica, see Fig.~\ref{fig3}(legend). Before the riddling bifurcation, each chimera replica inherits large neighborhoods for their basin of attraction, while the basin boundaries have fractal structure, $\mu=2.45$. After the riddling bifurcation, the basins become riddled with points belonging to the basins of the other replicas, $\mu=2.2$. Further decrease of coupling strength, at $\mu=1.9$ the basins are completely riddled and are called intermingled basins \cite{maistrenko1998transverse}. 

\section{\label{bl} Blowout Bifurcation Scenario}

The second scenario of chaotic switching involves blowout bifurcation of the standard chimera state. In a blowout bifurcation the attracting chaotic set in the invariant manifold loses transversal stability and becomes repulsive \cite{ott1994blowout}, when the transversal Lyapunov exponent becomes positive. Unlike the riddling bifurcation where it involves loss of transversal stability of a low-period orbit, the blowout bifurcation is characterized by the consecutive destabilization \cite{ashwin1994bubbling} of an infinite number of unstable orbits inside the chaotic manifold including high-period orbits \cite{nagai1997characterization,nagai1997periodic,yanchuk2003synchronization}.
A typical scenario of the blowout-provoked switching is illustrated in Fig.~\ref{fig5} where the coupling strength is fixed at $\mu = 2.3$ and the phase-shift is varied from $\alpha = 1.588$ increasing to $\alpha=1.5895$. The in-manifold $\lambda_1$ and transversal $\lambda_{tr}$ Lyapunov exponents of the standard chimera are shown in Fig.~\ref{fig5}(a) with blowout bifurcation point at $\alpha_b = 1.589091$ where the transversal Lyapunov exponent becomes positive. Nevertheless, the chaotic switching is observed before the blowout bifurcation as illustrated by Fig.~\ref{fig5}(b,c). After the blowout bifurcation the chimera manifold is transversally unstable and the system trajectory is dominated by short period switches Fig.~\ref{fig5}(d).
\begin{figure}
\includegraphics[width=\columnwidth]{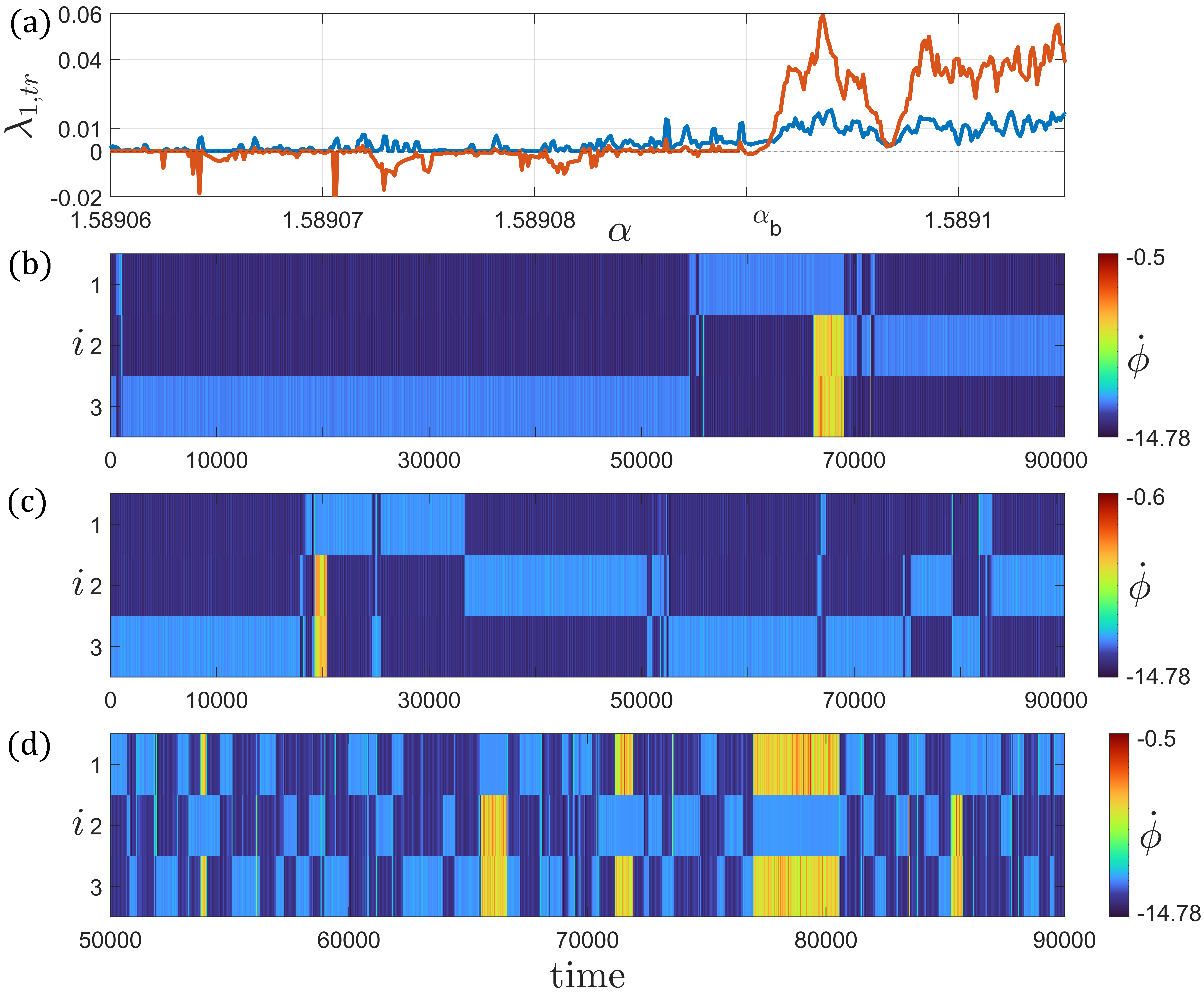}
\includegraphics[width=\columnwidth]{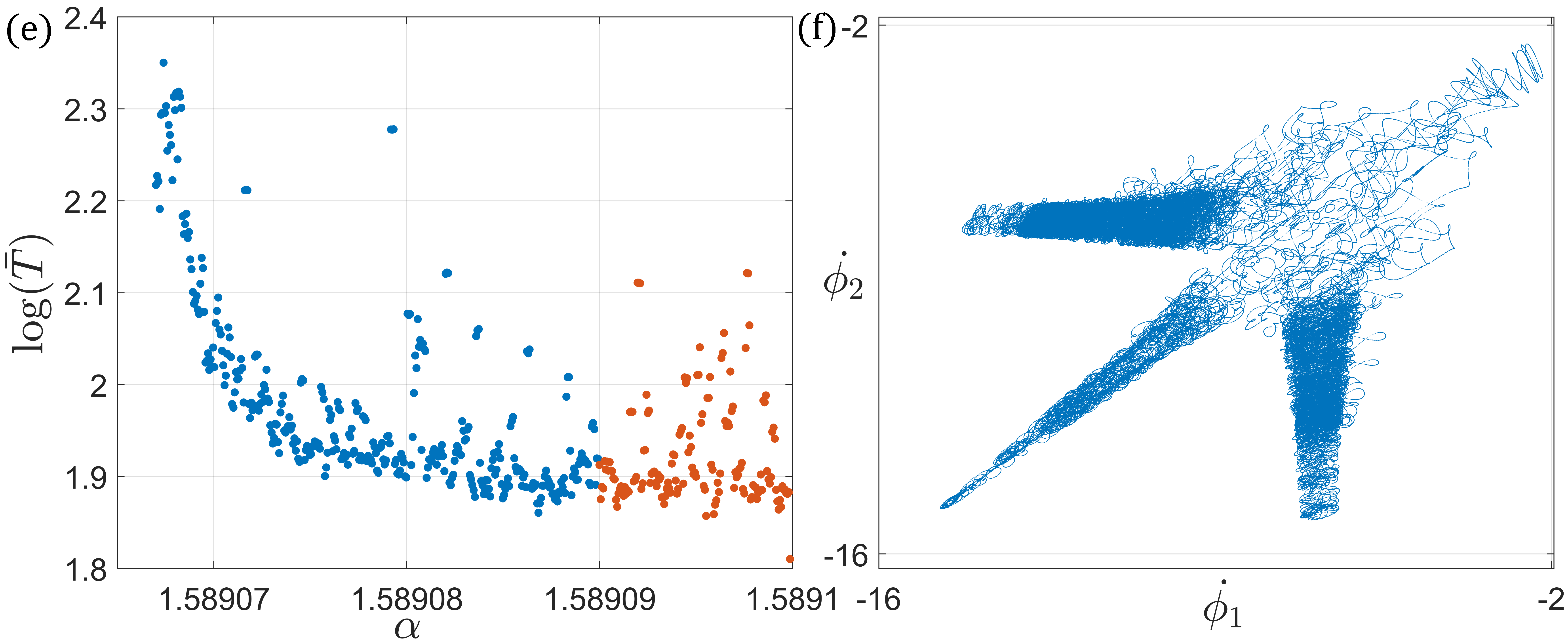}
\caption{\label{fig5} Blowout bifurcation scenario of chaotic switching. (a) Lyapunov exponents of the standard chimera state are calculated for fixed coupling strength $\mu = 2.3$ and varying phase-shift: the standard chimera is chaotic with one positive Lyapunov exponent. The switching dynamics of the standard chimera begins with riddling followed by consecutive transversal destabilization of infinite number of unstable periodic orbits. Switching examples for (b) $\alpha_1 = 1.589067$ and (b) $\alpha_2 = 1.589070$. The standard chimera then undergoes a blowout bifurcation at $\alpha_b = 1.589091$, where the transversal Lyapunov exponent of the chimera manifold becomes positive. (d) switching example after the blowout bifurcation $\alpha_3 = 1.5891$. Due to coexistence of the standard and inverted chimeras for these parameters, system trajectory can include switching between these two chimera types.  (e) the average switching period of between the standard chimeras plotted against phase-shift $\alpha$. (f) phase-velocity portrait of the three standard chimeras.}
\end{figure}
As shown in Fig.~\ref{fig5}(a), the standard chimera is chaotic with one positive Lyapunov exponent of order $\lambda_1 \approx 0.01$ and no switching is observed. The distinct characteristic of the blowout bifurcation is riddling transition just before the bifurcation, where transversal Lyapunov exponent  $\lambda_{tr}$ is still negative. As a result, the first switching in our simulation is observed at $\alpha_1 = 1.589067$, where $\lambda_{tr} = -0.011$. We note that the subspace of the standard chimera is not invariant anymore and the system trajectory occasionally leaves this subspace visiting the inverted chimeras before falling back to the standard chimera subspace, see the yellow band in Fig.~\ref{fig5}(b) in the time interval $[68000, 70000]$. With increase of $\alpha$ the more periodic orbits lose transversal stability\cite{ashwin1994bubbling,nagai1997characterization}, switching period decreases on average and more switches appear at the considered time interval, Fig.~\ref{fig5}(c). Finally, enough orbits have lost transversal stability and the chaotic manifold of the standard chimera becomes transversally unstable on average at the blowout bifurcation $\alpha=\alpha_b$ becoming a chaotic saddle. For phase-shift beyond the blowout $\alpha > \alpha_b$, the switching periods are very short and the system trajectory remains in the subspace of the standard and inverted chimeras, Fig.~\ref{fig5}(d).
The average switching period between the standard chimera replicas are presented in Fig.~\ref{fig5}(e). Increasing the phase-shift $\alpha$ the switching periods decrease much faster than that of the riddling bifurcation scenario, blue points in Fig.~\ref{fig5}(e). After the blowout bifurcation at $\alpha_b$, the switching periods are much shorter, but the change of the average switching period with respect to $\alpha$ does not vary significantly. We note that, due to the coexistence of the standard and inverted chimera types, the data is "littered" by the switching periods when the trajectory visits the inverted chimeras. Fig.~\ref{fig5}(f) shows the phase-velocity profile of the three standard chimera replicas at parameters $\alpha = 1.5891$ and $\mu = 2.3$. 

\section{\label{lt} Laminar-Turbulent Scenario}

The third scenario of chaotic switching involves the out-of-phase mixed-mode chimera no.2 as laminar parts embedded within a turbulent-like attractor. A similar scenario of switching between laminar chaos within a turbulent attractor has been recently observed in a shell model of fluid dynamics \cite{kato2024laminar}. In our scenario, the turbulence is understood when the fluctuations of the velocities dominate the dynamics in contrast to the chimeric behavior. Furthermore, the in-phase mixed-mode chimera no.2 coexists with the out-of-phase chimera and with the turbulence. The dynamics of the system then includes transient chaos \cite{kantz1985repellers, hsu1988strange} comprised of switching between the out-of-phase chimeras and the turbulence until the trajectory escapes to the stable in-phase chimera. 
The transient chaos behavior is typically due to the coexistence of chaotic saddles with globally nonattracting invariant chaotic sets. To this end, any trajectory with initial conditions in the neighborhood of the saddle set will eventually escape, hence transient chaos has a finite lifetime. However, since the nonattracting saddle set is invariant, initial conditions originated on the set will never leave, resulting in infinitely long transients. On the other hand, the nonattracting set has a zero Lebesgue measure of the basin and the infinitely long transients cannot be observed with random initial conditions \cite{lai2011transient}.
\begin{figure}
\includegraphics[width=\columnwidth]{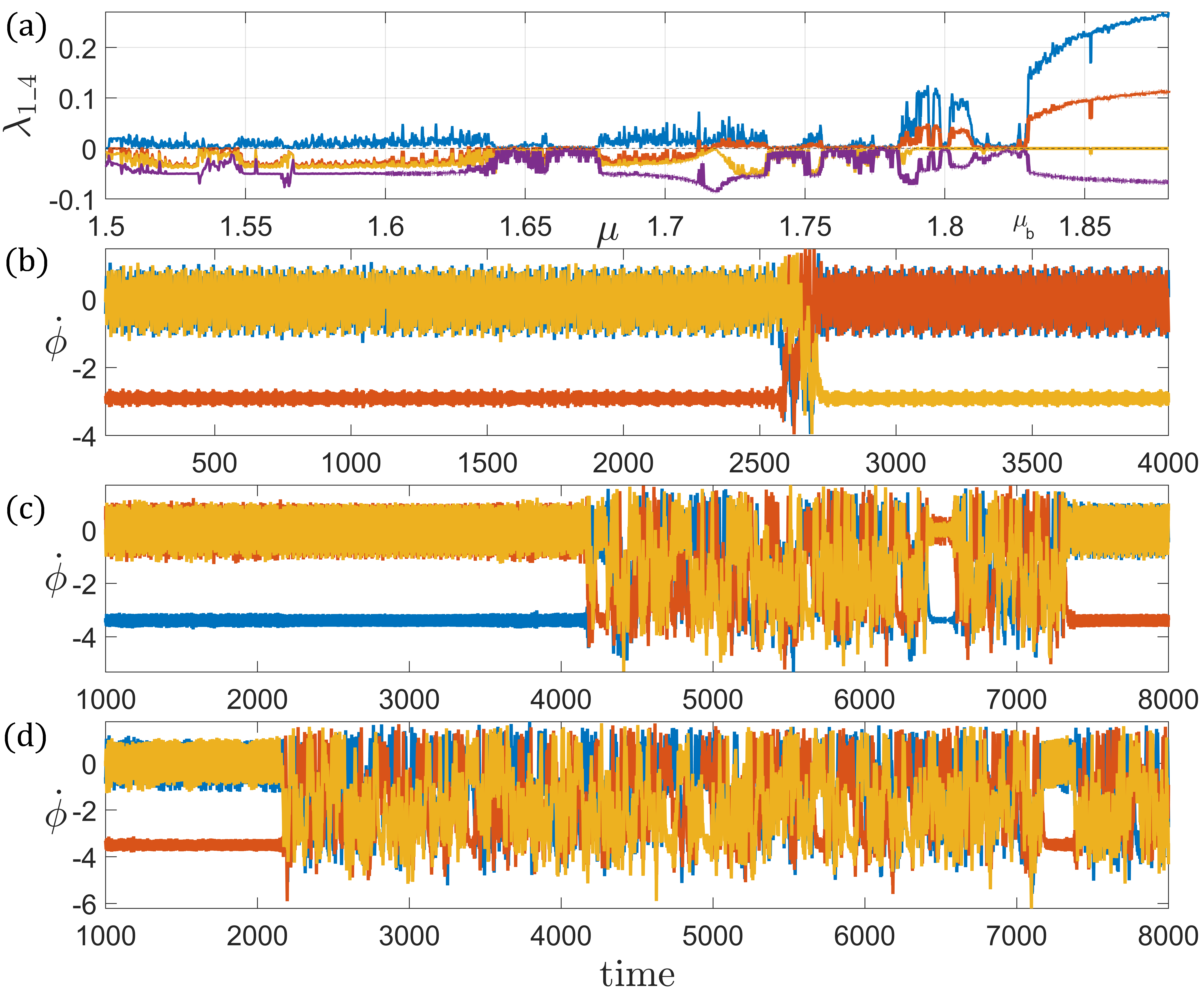}
\caption{\label{fig10} Laminar-Turbulent scenario of chaotic switching. (a) Lyapunov exponents of the out-of-phase mixed-mode chimera no.2 are calculated for fixed phase-shift $\alpha = 2.6$ and varying coupling strength: the chimera is chaotic with one positive Lyapunov exponent of the order $\lambda_{1, max} \approx 0.01$. In contrast to the previous two scenarios, the switching dynamics occur with the turbulence such that with increase of the coupling strength, the average period of the turbulence increases. Examples are shown for (b) $\mu_1 = 1.5067$, (c) $\mu_2 = 1.79$ and (d) $\mu_3 = 1.85$. We note that, the out-of -phase mixed-mode chimera no.2 undergoes a blowout bifurcation at $\mu_b = 1.82$ where the transversal Lyapunov exponent of the chimera manifold becomes positive.}
\end{figure}
\begin{figure*}
\includegraphics[width=\textwidth]{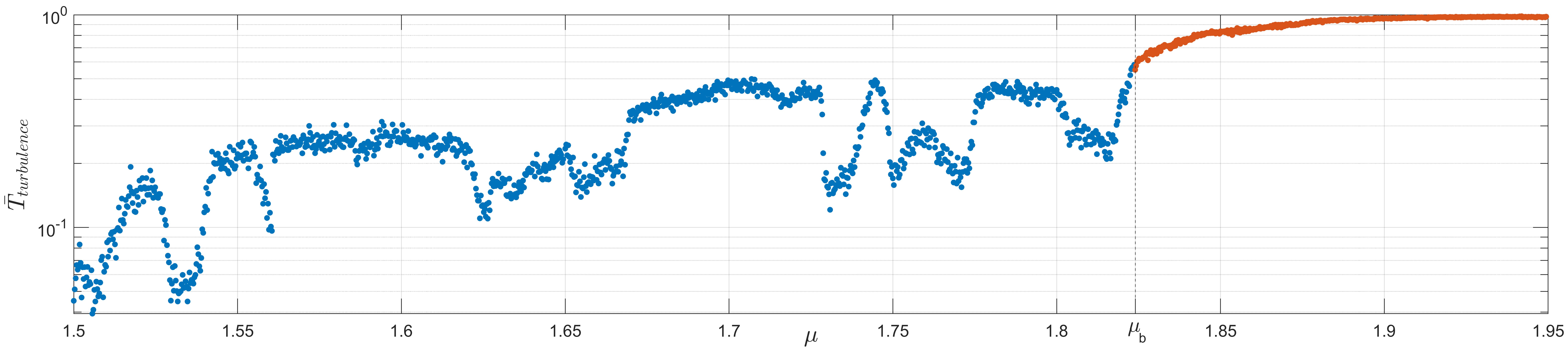}
\caption{\label{fig11} Average duration of the system trajectory trapped in the turbulence normalized to the total simulation time. The turbulence duration grows on average with increase of the coupling strength $\mu$ in a complicated manner, such that there are "dips" where the turbulence duration suddenly decreases. The dip regions are due to the earlier termination of the switching dynamics where the trajectory falls into the in-phase mixed-mode chimera no.2. After the blowout bifurcation of the out-of-phase mixed-mode chimera no.2, system trajectory is mostly in the turbulent attractor.}
\end{figure*}

A typical scenario of the laminar-turbulent switching is illustrated in Fig.~\ref{fig10} where the phase-shift is fixed at $\alpha = 2.6$ and coupling strength is varied from $\mu = 1.5$ increasing to $\mu = 1.9$. Four of the Lyapunov exponents of the out-of-phase mixed-mode chimera no.2 are shown in Fig.~\ref{fig10}(a).
Initially, the out-of-phase chimera is a chaotic saddle with one positive Lyapunov exponent of the order $\lambda_{1,max} \sim 0.01$. On the other hand, the out-of-phase chimera is \textit{transversally weakly stable} \cite{kapitaniak1999blowout} in the Milnor sense, meaning there exists a non-zero measure set of trajectories originated in the vicinity of the out-of-phase chimera that are attracted to it. There exist some trajectories, however, that escape the out-of-phase chimera. These trajectories then are trapped in the turbulence for some time before returning to the out-of-phase chimera, Fig.~\ref{fig10}(b) at $\mu = 1.5067$. The system trajectory, however, will eventually escape the chaotic switching, falling into the in-phase chimera. Increasing the coupling strength to $\mu = 1.79$, Fig.~\ref{fig10}(c), the number of switches and the duration of turbulent state increase until the chaotic saddle of the out-of phase chimera goes through a blowout bifurcation at $\mu_b = 1.82$, where the full measure of trajectories in the neighborhood of the chaotic saddle are repelled away from it. After the blowout bifurcation, the trajectory mostly stays in the turbulence, Fig.~\ref{fig10}(d) at $\mu = 1.85$. We note that, a characteristic of the transient chaos is the sensitivity of the transient lifetime to the initial conditions: trajectories originated from nearby initial conditions may have drastically different lifetimes. For this chaotic switching scenario between the out-of-phase chimeras and the turbulence, not only the lifetime of the transient, but the switching sequences between the out-of-phase chimeras, their period and duration of the trajectory in the turbulence are extremely sensitive to the initial conditions as well.
\begin{figure}
\includegraphics[width=\columnwidth]{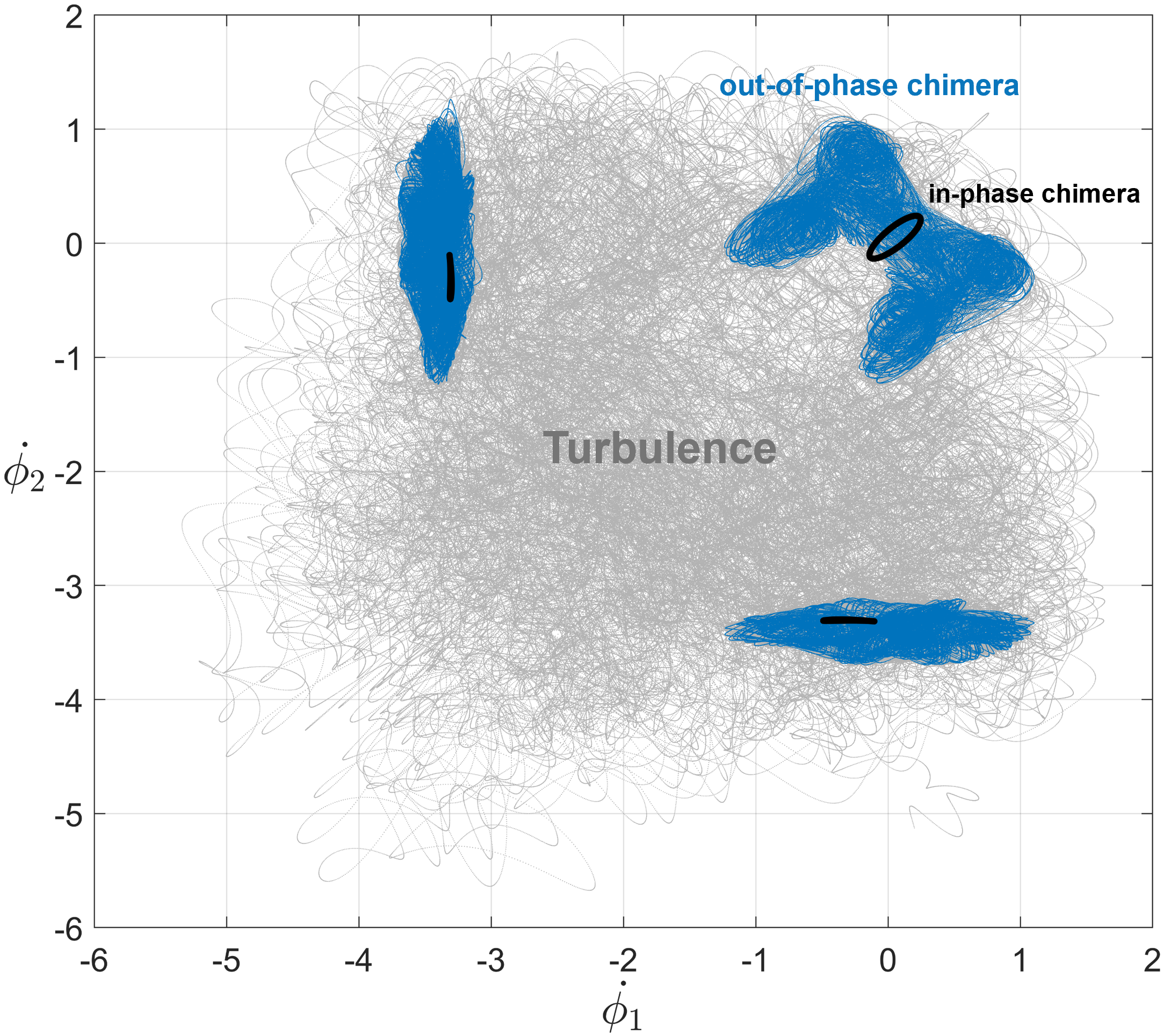}
\includegraphics[width=\columnwidth]{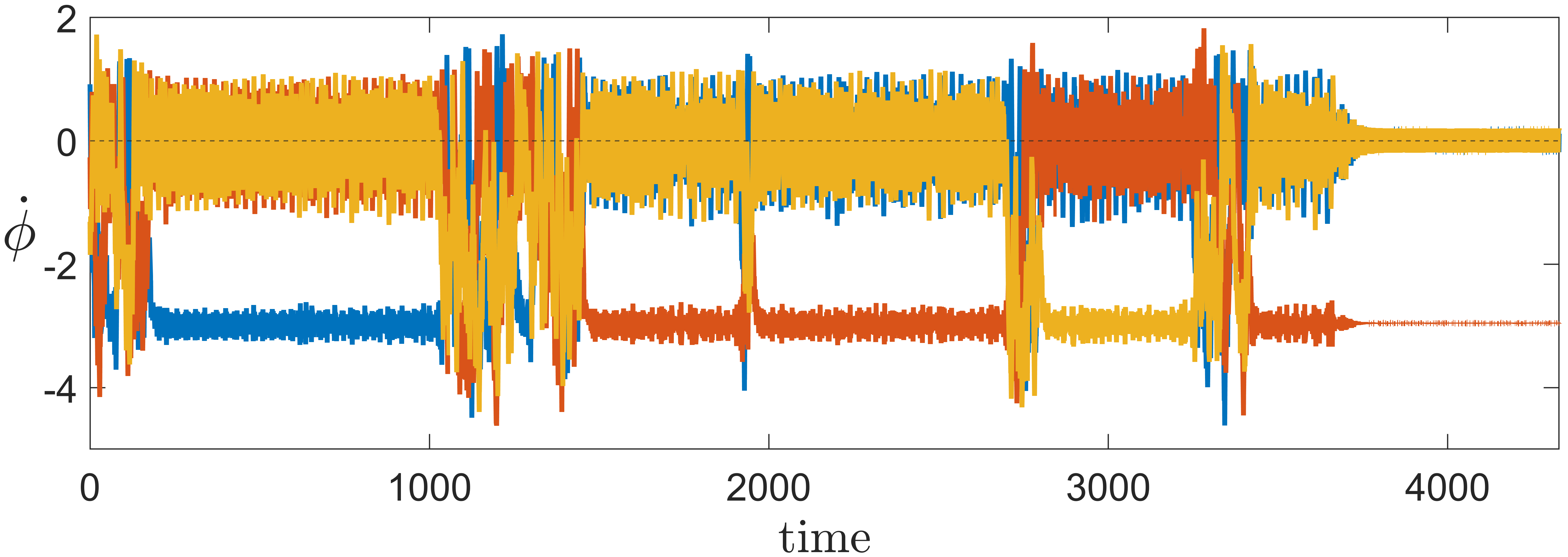}
\caption{\label{fig9} (a) Phase-velocity portrait of the system trajectory including in-phase mixed-mode chimera no.2 (black), out-of-phase mixed-mode chimera no.2 (blue) and turbulent-like behavior (gray). (b) The velocity plot of the switching dynamics between chaotic out-of-phase chimera replicas and the turbulence, terminating at $t = 3800$ to the in-phase chimera. The coupling parameters are $(\alpha, \mu)$ = (2.6, 1.55).}
\end{figure}

Fig.~\ref{fig11} shows the total time of the trajectory trapped in the turbulence normalized to the simulation time. The figure is obtained by varying coupling strength with step size $\Delta \mu = 0.003$; each data point is obtained by averaging of the turbulence duration over 100 iterations where the initial conditions are randomized around the out-of-phase chimera with Gaussian distribution $\mathcal{N}(0, 0.1)$. In the beginning, for small values of coupling strength $\mu$, the duration of the turbulence is not significant. With the increase of $\mu$, the duration of the turbulence increases on average. Before the blowout bifurcation of the out-of-phase chimera for $\mu<\mu_b$, the trajectory consists of switchings between the out-of-phase chimera and turbulence until falling into the regular in-phase chimera. The sudden drops of the turbulence duration in Fig.~\ref{fig11} correspond to the regions where the trajectory was attracted earlier to the in-phase chimera. After the blowout bifurcation at $\mu > \mu_b$, however, the trajectory mostly consists of turbulence and rarely falls into the in-phase chimera.

The phase-velocity portrait of and a typical system trajectory of this scenario is shown in Fig.~\ref{fig9}. The blue segments represent the out-of-phase chimeras, the black circles inside them are the in-phase chimeras, while the gray part is the turbulence. In contrast to the previous scenarios, the switches between the chimeras are not instantaneous such that the trajectory, after leaving the chimera, is trapped in the turbulence for some time. For the system parameters where the turbulence coexists with asynchronous fixed points, we observe transient turbulent behavior until the trajectory escapes to the asynchronous fixed point.

\section{\label{csd} Conclusion and Discussion} 

In this paper, we reported on three scenarios for the chaotic switching behavior between chimera states in a small network of three globally coupled pendula. 
Appearance of the chimera states in pendula networks was previously reported \cite{ebrahimzadeh2022mixed} in the attractive coupling regime $\alpha < \pi / 2$. In the repulsive coupling regime $\alpha > \pi / 2$ we found that the chimeras become chaotic and hyperchaotic in some regions of the parameter plane $(\alpha, \mu)$. Switching states arise as the chaotic chimeras lose stability in transverse direction with respect to their invariant manifolds, becoming chaotic saddles. Three scenarios of chaotic switching are found where chaotic chimeras lose transverse stability in a riddling bifurcation, a blowout bifurcation and a laminar-turbulent scenario. The chaotic switching of the laminar-turbulent scenario is transient such that the system trajectory eventually escapes the switching behavior and ends in a regular attractor (fixed point or limit cycle) that coexists with it. The lifetime of the transient chaotic switching, the sequence of switches and their period are extremely sensitive to the initial conditions, which supplies an additional level of complexity to the system behavior, making it even more unpredictable.

Our analysis suggests that the robust mixed-mode chimeric behavior and chaotic switching states can generically describe the complex dynamics of diverse pendula-like systems, ranging from spiking neural networks, power grids to neuromorphic systems. It has been shown that, analog dynamical systems can be used in solving different class' of NP-hard problems \cite{molnar2018continuous} where the solution of the NP-hard problem is a solution of the dynamical system. Strikingly, the hardness of the problem correlates with emergence of chaotic saddles that trap the trajectory for a transient time \cite{ercsey2011optimization,varga2016order}. With technological advancement of novel material with neuromorphic capabilities, complex dynamical systems such as oscillator networks are emerging candidates for neuromorphic hardware. To that end, our analysis suggests deeper understanding of the collective system dynamics is needed for bio-inspired problem solving and computation.

\appendix

\section{phenomenology}
\label{A1}

The system~(\ref{PN}) have been simulated by implementing a standard 4th order Runge-Kutta integrator with constant time step $\Delta t = 0.01$. The stability regions of the dynamical behavior of system~(\ref{PN}), sans chaotic switching states, in the $(\alpha, \mu)$ plane are obtained using continuation method with step sizes $\Delta \alpha = 0.01$ and $\Delta \mu = 0.01$. Each data point in $(\alpha, \mu)$ plane is simulated for $T = 500$ with disregarded transient time $T_{transient} = 300$ and noise intensity $10^{-14}$. The chaotic switching states are obtained with fixed random initial conditions with simulation time $T = 2000$ with disregarded transient time $T_{transient} = 200$, step sizes $\Delta \alpha = 0.005$ and $\Delta \mu = 0.01$ and no noise present. We identify the chimera clusters with \texttt{clusterdata} function in MATLAB, with the average frequency distance \texttt{cutoff}$=0.1$.
The Lyapunov exponents and Kaplan-Yorke dimension are obtained using \texttt{DynamicalSystems.jl}\cite{Datseris2018} and \texttt{ChaosTools.jl} with calculation over 500,000 evolution steps.

\section{transition to chaos}
\label{A2}

\begin{figure}
\includegraphics[width=\columnwidth]{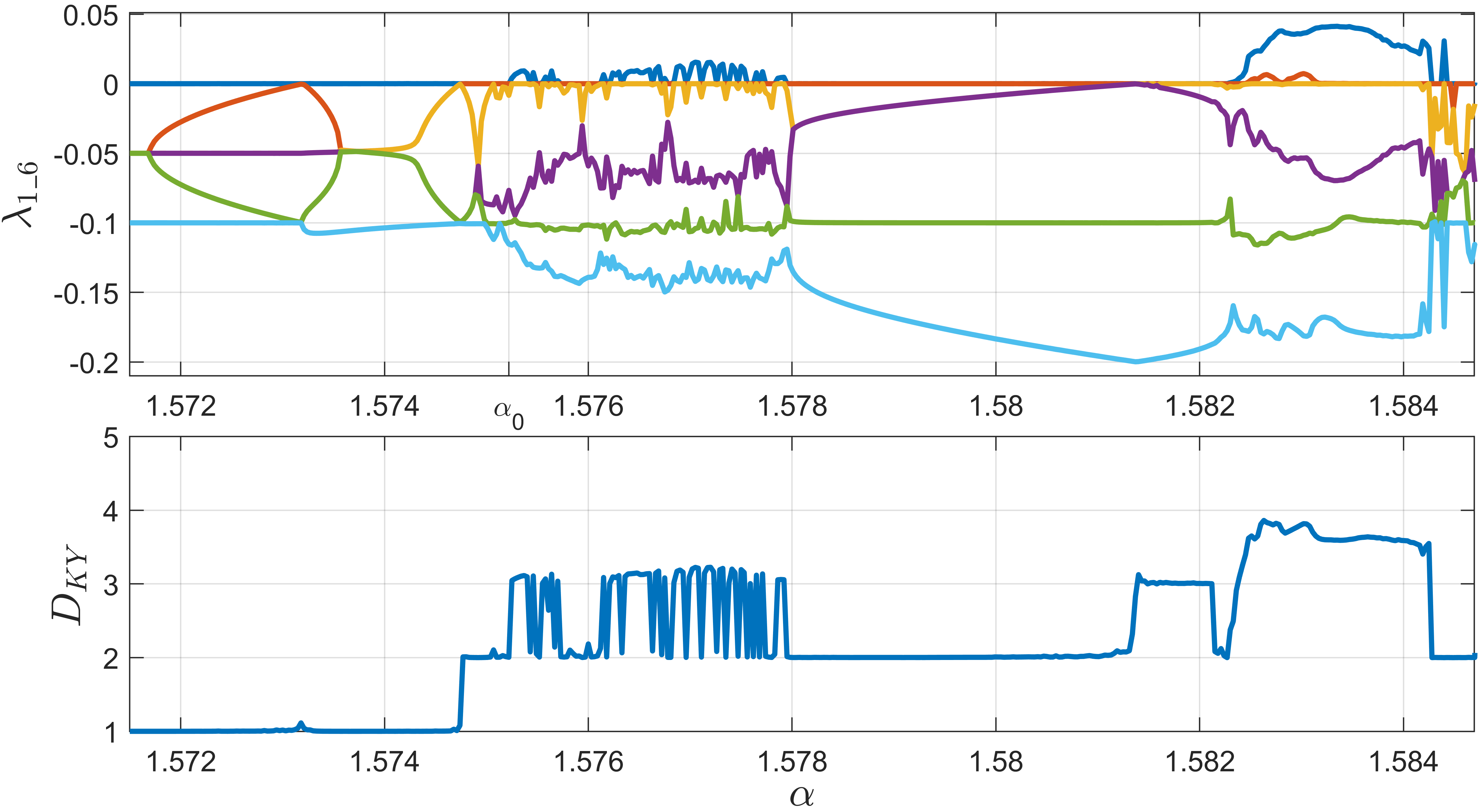}
\caption{\label{fig12} Lyapunov exponents $\lambda_{1\_6}$ and Kaplan-Yorke dimension $D_{KY}$ of the standard chimera at $\mu = 2.5$.}
\end{figure}
\begin{figure}
\includegraphics[width=\columnwidth]{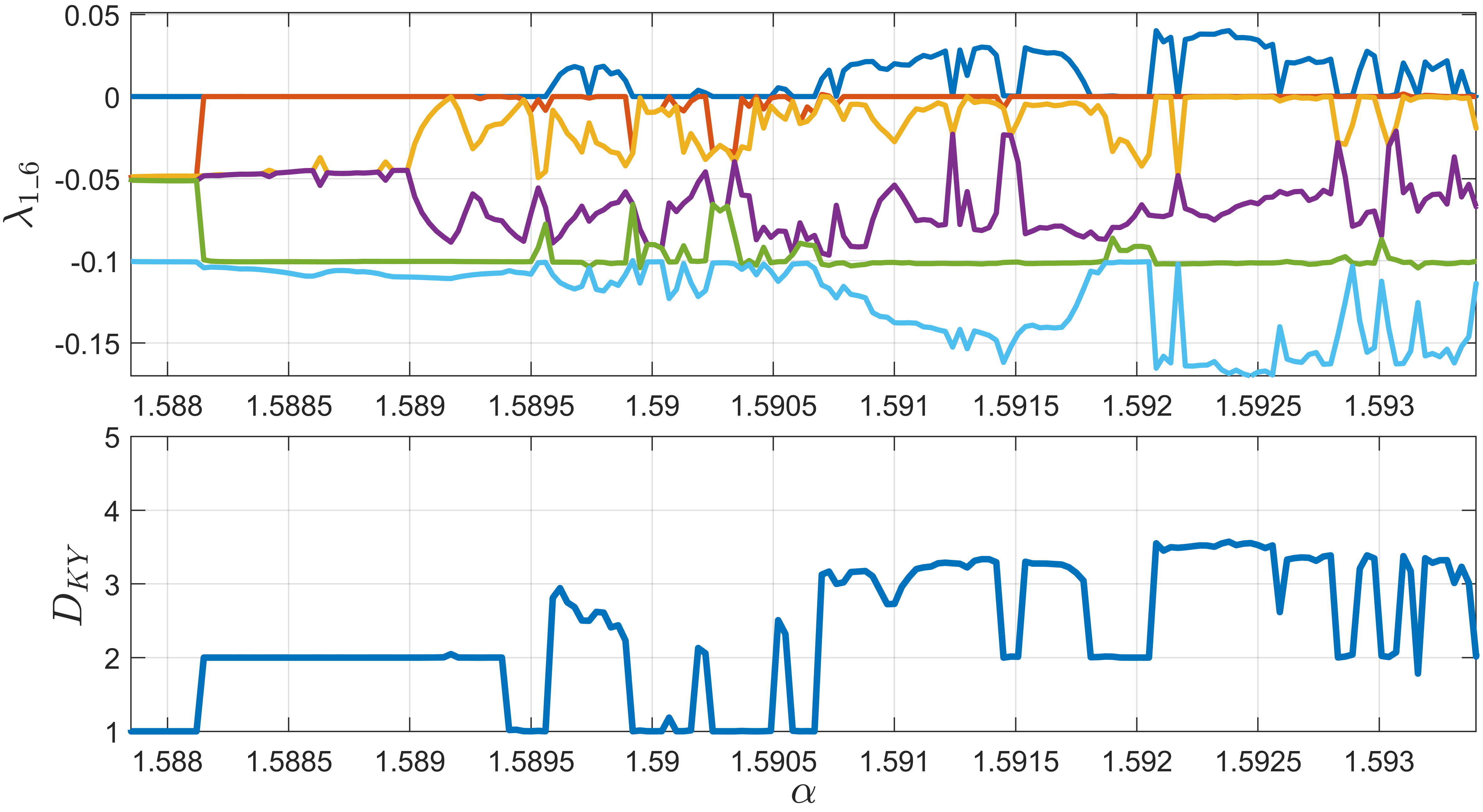}
\caption{\label{fig13}  Lyapunov exponents $\lambda_{1\_6}$ and Kaplan-Yorke dimension $D_{KY}$ of the inverted chimera at $\mu = 2.5$.}
\end{figure}
The evolution of the standard and inverted chimera states are demonstrated in Fig.~\ref{fig12} with their Lyapunov exponents (top) and Kaplan-Yorke dimension (bottom).  Varying the phase-shift $\alpha$ and fixing the coupling strength $\mu=2.5$, the standard chimera state is a periodic attractor with largest Lyapunov exponent zero $\lambda_1 = 0$ and dimension one $D_{KY} = 1$. Increasing $\alpha \approx 1.575$, the periodic standard chimera states becomes quasi-periodic with its dynamics lying on a two dimensional torus $D_{KY} = 2$. The first chaotic dynamics are observed with the largest Lyapunov exponent  becoming positive $\lambda_1 \approx 0.0017$, between fractal windows of stability, Fig.~\ref{fig12}(top) . The chaotic dynamic of the standard chimera state lies in a three dimensional torus $D_{KY} = 3$ with distinct configuration. After another alteration between two and three dimensional tori, the standard chimera torus increases in dimension near four $D_{KY} \sim 4$. The inverted chimera follows the same scenario of transition to chaos, Fig.~\ref{fig13}.

\nocite{*}
\bibliography{Refs}

@PREAMBLE{
 "\providecommand{\noopsort}[1]{}" 
 # "\providecommand{\singleletter}[1]{#1}%" 
}

@article{kuramoto2002coexistence,
  title={Coexistence of coherence and incoherence in nonlocally coupled phase oscillators},
  author={Kuramoto, Yoshiki and Battogtokh, Dorjsuren},
  journal={Nonlinear Phenom. Complex Syst. 5, 380–385},
  year={2002}
}

@article{PhysRevLett.93.174102,
  title = {Chimera States for Coupled Oscillators},
  author = {Abrams, Daniel M. and Strogatz, Steven H.},
  journal = {Phys. Rev. Lett.},
  volume = {93},
  issue = {17},
  pages = {174102},
  numpages = {4},
  year = {2004},
  month = {Oct},
  publisher = {American Physical Society},
  doi = {10.1103/PhysRevLett.93.174102},
  url = {https://link.aps.org/doi/10.1103/PhysRevLett.93.174102}
}

@article{davidsen2024introduction,
  title={Introduction to Focus Issue: Chimera states: From theory and experiments to technology and living systems},
  author={Davidsen, J{\"o}rn and Maistrenko, Yuri and Showalter, Kenneth},
  journal={Chaos: An Interdisciplinary Journal of Nonlinear Science},
  volume={34},
  number={12},
  year={2024},
  publisher={AIP Publishing}
}

@article{Panaggio_2015,
	doi = {10.1088/0951-7715/28/3/r67},
	url = {https://doi.org/10.1088/0951-7715/28/3/r67},
	year = 2015,
	month = {feb},
	publisher = {{IOP} Publishing},
	volume = {28},
	number = {3},
	pages = {R67--R87},
	author = {Mark J Panaggio and Daniel M Abrams},
	title = {Chimera states: coexistence of coherence and incoherence in networks of coupled oscillators},
	journal = {Nonlinearity},
}

@article{scholl2016synchronization,
  title={Synchronization patterns and chimera states in complex networks: Interplay of topology and dynamics},
  author={Sch{\"o}ll, Eur},
  journal={The European Physical Journal Special Topics},
  volume={225},
  number={6},
  pages={891--919},
  year={2016},
  publisher={Springer}
}

@article{omel2019chimerapedia,
  title={Chimerapedia: coherence--incoherence patterns in one, two and three dimensions},
  author={Omel’chenko, E and Knobloch, Edgar},
  journal={New Journal of Physics},
  volume={21},
  number={9},
  pages={093034},
  year={2019},
  publisher={IOP Publishing}
}

@book{zakharova2020chimera,
  title={Chimera Patterns in Networks},
  author={Zakharova, Anna},
  year={2020},
  publisher={Springer}
}

@article{mishra2023chimeras,
  title={Chimeras in globally coupled oscillators: A review},
  author={Mishra, Arindam and Saha, Suman and Dana, Syamal K},
  journal={Chaos: An Interdisciplinary Journal of Nonlinear Science},
  volume={33},
  number={9},
  year={2023},
  publisher={AIP Publishing}
}

@article{ashwin2015weak,
  title={Weak chimeras in minimal networks of coupled phase oscillators},
  author={Ashwin, Peter and Burylko, Oleksandr},
  journal={Chaos: An Interdisciplinary Journal of Nonlinear Science},
  volume={25},
  number={1},
  pages={013106},
  year={2015},
  publisher={AIP Publishing LLC}
}

@article{jaros2018solitary,
  title={Solitary states for coupled oscillators with inertia},
  author={Jaros, Patrycja and Brezetsky, Serhiy and Levchenko, Roman and Dudkowski, Dawid and Kapitaniak, Tomasz and Maistrenko, Yuri},
  journal={Chaos: An Interdisciplinary Journal of Nonlinear Science},
  volume={28},
  number={1},
  pages={011103},
  year={2018},
  publisher={AIP Publishing LLC}
}

@article{maistrenko2017smallest,
  title={Smallest chimera states},
  author={Maistrenko, Yuri and Brezetsky, Serhiy and Jaros, Patrycja and Levchenko, Roman and Kapitaniak, Tomasz},
  journal={Physical Review E},
  volume={95},
  number={1},
  pages={010203},
  year={2017},
  publisher={APS}
}

@article{tricomi1933integrazione,
  title={Integrazione di un'equazione differenziale presentatasi in elettrotecnica},
  author={Tricomi, Francesco},
  journal={Annali della Scuola Normale Superiore di Pisa-Scienze Fisiche e Matematiche},
  volume={2},
  number={1},
  pages={1--20},
  year={1933}
}

@article{ebrahimzadeh2022mixed,
  title={Mixed-mode chimera states in pendula networks},
  author={Ebrahimzadeh, P and Schiek, M and Maistrenko, Y},
  journal={Chaos: An Interdisciplinary Journal of Nonlinear Science},
  volume={32},
  number={10},
  year={2022},
  publisher={AIP Publishing}
}

@article{brezetsky2021chimera,
  title={Chimera complexity},
  author={Brezetsky, Serhiy and Jaros, Patrycja and Levchenko, Roman and Kapitaniak, Tomasz and Maistrenko, Yuri},
  journal={Physical Review E},
  volume={103},
  number={5},
  pages={L050204},
  year={2021},
  publisher={APS}
}

@article{goldschmidt2019blinking,
  title={Blinking chimeras in globally coupled rotators},
  author={Goldschmidt, Richard Janis and Pikovsky, Arkady and Politi, Antonio},
  journal={Chaos: An Interdisciplinary Journal of Nonlinear Science},
  volume={29},
  number={7},
  year={2019},
  publisher={AIP Publishing}
}

@article{zhang2020critical,
  title={Critical switching in globally attractive chimeras},
  author={Zhang, Yuanzhao and Nicolaou, Zachary G and Hart, Joseph D and Roy, Rajarshi and Motter, Adilson E},
  journal={Physical Review X},
  volume={10},
  number={1},
  pages={011044},
  year={2020},
  publisher={APS}
}

@article{rabinovich2001dynamical,
  title={Dynamical encoding by networks of competing neuron groups: winnerless competition},
  author={Rabinovich, M and Volkovskii, A and Lecanda, P and Huerta, Ram{\'o}n and Abarbanel, Henry DI and Laurent, Gilles},
  journal={Physical review letters},
  volume={87},
  number={6},
  pages={068102},
  year={2001},
  publisher={APS}
}

@article{ott1994blowout,
  title={Blowout bifurcations: the occurrence of riddled basins and on-off intermittency},
  author={Ott, Edward and Sommerer, John C},
  journal={Physics Letters A},
  volume={188},
  number={1},
  pages={39--47},
  year={1994},
  publisher={Elsevier}
}

@article{kraut2002multistability,
  title={Multistability, noise, and attractor hopping: The crucial role of chaotic saddles},
  author={Kraut, Suso and Feudel, Ulrike},
  journal={Physical Review E},
  volume={66},
  number={1},
  pages={015207},
  year={2002},
  publisher={APS}
}

@article{ansmann2016self,
  title={Self-induced switchings between multiple space-time patterns on complex networks of excitable units},
  author={Ansmann, Gerrit and Lehnertz, Klaus and Feudel, Ulrike},
  journal={Physical Review X},
  volume={6},
  number={1},
  pages={011030},
  year={2016},
  publisher={APS}
}

@article{lai1996riddling,
  title={Riddling bifurcation in chaotic dynamical systems},
  author={Lai, Ying-Cheng and Grebogi, Celso and Yorke, James A and Venkataramani, SC},
  journal={Physical review letters},
  volume={77},
  number={1},
  pages={55},
  year={1996},
  publisher={APS}
}

@article{kato2024laminar,
  title={Laminar chaotic saddle within a turbulent attractor},
  author={Kato, Hibiki and Kobayashi, Miki U and Saiki, Yoshitaka and Yorke, James A},
  journal={Physical Review E},
  volume={110},
  number={5},
  pages={L052202},
  year={2024},
  publisher={APS}
}

@book{lai2011transient,
  title={Transient chaos: complex dynamics on finite time scales},
  author={Lai, Ying-Cheng and T{\'e}l, Tam{\'a}s},
  volume={173},
  year={2011},
  publisher={Springer Science \& Business Media}
}

@article{kantz1985repellers,
  title={Repellers, semi-attractors, and long-lived chaotic transients},
  author={Kantz, Holger and Grassberger, Peter},
  journal={Physica D: Nonlinear Phenomena},
  volume={17},
  number={1},
  pages={75--86},
  year={1985},
  publisher={Elsevier}
}

@article{hsu1988strange,
  title={Strange saddles and the dimensions of their invariant manifolds},
  author={Hsu, Guan-Hsong and Ott, Edward and Grebogi, Celso},
  journal={Physics Letters A},
  volume={127},
  number={4},
  pages={199--204},
  year={1988},
  publisher={Elsevier}
}

@article{kapitaniak1999blowout,
  title={Blowout bifurcation of chaotic saddles},
  author={Kapitaniak, Tomasz and Lai, Ying-Cheng and Grebogi, Celso},
  journal={Discrete Dynamics in Nature and Society},
  volume={3},
  number={1},
  pages={9--13},
  year={1999},
  publisher={Wiley Online Library}
}

@article{ashwin2005instability,
  title={When instability makes sense},
  author={Ashwin, Peter and Timme, Marc},
  journal={Nature},
  volume={436},
  number={7047},
  pages={36--37},
  year={2005},
  publisher={Nature Publishing Group UK London}
}

@article{ercsey2011optimization,
  title={Optimization hardness as transient chaos in an analog approach to constraint satisfaction},
  author={Ercsey-Ravasz, M{\'a}ria and Toroczkai, Zolt{\'a}n},
  journal={Nature Physics},
  volume={7},
  number={12},
  pages={966--970},
  year={2011},
  publisher={Nature Publishing Group UK London}
}

@article{varga2016order,
  title={Order-to-chaos transition in the hardness of random Boolean satisfiability problems},
  author={Varga, Melinda and Sumi, R{\'o}bert and Toroczkai, Zolt{\'a}n and Ercsey-Ravasz, M{\'a}ria},
  journal={Physical Review E},
  volume={93},
  number={5},
  pages={052211},
  year={2016},
  publisher={APS}
}

@article{rossi2025transients,
  title={Transients versus network interactions give rise to multistability through trapping mechanism},
  author={Rossi, Kalel L and Medeiros, Everton S and Ashwin, Peter and Feudel, Ulrike},
  journal={Chaos: An Interdisciplinary Journal of Nonlinear Science},
  volume={35},
  number={3},
  year={2025},
  publisher={AIP Publishing}
}

@article{molnar2018continuous,
  title={A continuous-time MaxSAT solver with high analog performance},
  author={Moln{\'a}r, Botond and Moln{\'a}r, Ferenc and Varga, Melinda and Toroczkai, Zolt{\'a}n and Ercsey-Ravasz, M{\'a}ria},
  journal={Nature communications},
  volume={9},
  number={1},
  pages={4864},
  year={2018},
  publisher={Nature Publishing Group UK London}
}

@article{alexander1992riddled,
  title={Riddled basins},
  author={Alexander, JC and Yorke, James A and You, Zhiping and Kan, Ittai},
  journal={International Journal of Bifurcation and Chaos},
  volume={2},
  number={04},
  pages={795--813},
  year={1992},
  publisher={World Scientific}
}

@article{maistrenko1998transverse,
  title={Transverse instability and riddled basins in a system of two coupled logistic maps},
  author={Maistrenko, Yu L and Maistrenko, VL and Popovich, A and Mosekilde, Erik},
  journal={Physical Review E},
  volume={57},
  number={3},
  pages={2713},
  year={1998},
  publisher={APS}
}

@article{Datseris2018,
 doi = {10.21105/joss.00598}, 
 year = {2018}, 
 publisher = {The Open Journal}, 
 volume = {3}, 
 number = {23}, 
 pages = {598}, 
 author = {Datseris, George},
 title = {DynamicalSystems.jl: A Julia software library for chaos and nonlinear dynamics},
 journal = {Journal of Open Source Software} 
}

@article{ashwin1994bubbling,
  title={Bubbling of attractors and synchronisation of chaotic oscillators},
  author={Ashwin, Peter and Buescu, Jorge and Stewart, Ian},
  journal={Physics Letters A},
  volume={193},
  number={2},
  pages={126--139},
  year={1994},
  publisher={Elsevier}
}

@article{nagai1997characterization,
  title={Characterization of blowout bifurcation by unstable periodic orbits},
  author={Nagai, Yoshihiko and Lai, Ying-Cheng},
  journal={Physical Review E},
  volume={55},
  number={2},
  pages={R1251},
  year={1997},
  publisher={APS}
}

@article{nagai1997periodic,
  title={Periodic-orbit theory of the blowout bifurcation},
  author={Nagai, Yoshihiko and Lai, Ying-Cheng},
  journal={Physical Review E},
  volume={56},
  number={4},
  pages={4031},
  year={1997},
  publisher={APS}
}

@article{yanchuk2003synchronization,
  title={Synchronization of time-continuous chaotic oscillators},
  author={Yanchuk, Sergiy and Maistrenko, Yuri and Mosekilde, Erik},
  journal={Chaos: An Interdisciplinary Journal of Nonlinear Science},
  volume={13},
  number={1},
  pages={388--400},
  year={2003},
  publisher={American Institute of Physics}
}

\end{document}